\documentclass{article}
\usepackage[utf8]{inputenc}

\title{\vspace{-1.5cm}Neural Optimization: Understanding Trade-offs\\ with Pareto Theory}
\author{Fabian Pallasdies, Philipp Norton, Jan-Hendrik Schleimer, Susanne Schreiber}

\date{%
\small
    Institute for Theoretical Biology, Department of Biology, Humboldt-Universität zu Berlin, Berlin, Germany\\
    Bernstein Center for Computational Neuroscience, Berlin, Germany\\
    \vspace{0.5cm}
    April 2021
}

\usepackage[square,numbers,sort&compress]{natbib}
\usepackage{graphicx}
\usepackage[svgnames]{xcolor}
\usepackage{float}
\usepackage[a4paper, total={6in, 9in}]{geometry}
\usepackage[doublespacing]{setspace}

\graphicspath{ {./figures/} }
\DeclareGraphicsExtensions{.pdf}

\begin{document}

\maketitle


\begin{abstract}

\noindent Nervous systems, like any organismal structure, have been shaped by evolutionary processes to increase fitness. The resulting neural 'bauplan' has to account for multiple objectives simultaneously, including computational function as well as additional factors like robustness to environmental changes and energetic limitations. Oftentimes these objectives compete and a quantification of the relative impact of individual optimization targets is non-trivial. \emph{Pareto optimality} offers a theoretical framework to decipher objectives and trade-offs between them. We, therefore, highlight Pareto theory as a useful tool for the analysis of neurobiological systems, from biophysically-detailed cells to large-scale network structures and behavior. The Pareto approach can help to assess optimality, identify relevant objectives and their respective impact, and formulate testable hypotheses.
\end{abstract}
\vspace{0.5cm}

\begin{center}
\begin{spacing}{1.3}
\fcolorbox{darkgray}{Gainsboro}{\begin{minipage}{\linewidth}
\textbf{\Large Highlights}
\begin{enumerate}
    \small
    \item Nervous systems are not only optimized for function but also robustness, efficiency, and flexibility. Depending on the ecological niche, more than one constraint leaves its imprint onto the neural structure.
    
    \item The concept of Pareto optimality allows us to measure the influence of individual constraints on nervous system optimization. Neural design can be optimal along so-called \emph{Pareto boundaries}, from electrical properties and cellular morphology to behavior.
    
    \item Pareto boundaries can be identified via parameter exploration in mathematical models or by extraction of correlations in experimental data that offer sufficient statistics on biological traits.
    
    \item The importance of the functional readout (e.g. information processed in sensory systems, robustness of motor output) determines the acceptable investment in other constraints.

\end{enumerate}
\end{minipage}}
\end{spacing}
\end{center}

\section*{Introduction}

Neural optimization is easy to spot when it gives rise to an exceptional function, like the processing of sounds on the microsecond scale when the owl catches its prey or the sensation of electromagnetic waves by electric fish. 
Although at first glance a specific function might seem to have dominated the optimization process, usually several tasks and additional constraints need to be fulfilled in parallel. Accordingly, robustness of performance has to be ensured despite variable temperatures, even when energetic supplies fade, or when anatomical space is scarce. The relative weighting between demands can dramatically alter which solution is favored, and it is often not self-evident which objectives have most influenced the genesis of a particular neuronal structure. As biological systems cannot be optimal at all tasks \cite{Shoval_Sheftel_Shinar_Hart_Ramote_Mayo_Dekel_Kavanagh_Alon_2012, Szekely_Korem_Moran_Mayo_Alon_2015}, trade-offs are ubiquitous. 

While some constraints like energetic limitations have been extensively reviewed \cite{Niven_2016,Yu_Yu_2017, Rittschof_Schirmeier_2018}, the literature on other neurobiological trade-offs is more scattered; even fewer attempts have been made to construct a general framework. Del Giudice \& Crespi \cite{DelGiudice_Crespi_2018} have studied four categories of functional properties in neuroscience: performance, efficiency, robustness and flexibility, and have investigated the trade-offs between them. They also recognized the need for a methodological framework in which interrelations between constraints can be analyzed and hypotheses can be formed and tested. Here, we provide a review of recent studies on trade-offs in neural systems with a particular focus on the concept of \emph{Pareto optimality} as a  useful methodology for research on neural optimization principles.

\section*{Optimality Theories with Single Objectives}

Many optimality theories of the nervous system have targeted single objectives, such as the maximization of information transfer \cite{Laughlin_1981, Sengupta_Stemmler_Friston_2013}, efficient representation \cite{Barlow_1961}, the minimization of wiring cost \cite{Wang_Clandinin_2016}, free energy \cite{Sengupta2016Gauge} and energy consumption \cite{Niven_Laughlin_2008, Sengup2010ActionPotentia, YuYu2012WarmBodyTempe, Niven_2016}. Each of these objectives is thought to monotonically impact organismal fitness.  
Singling out a specific objective as the predominant contributor to fitness results in so-called \emph{archetypes}, i.e., individuals that are optimal with respect to one objective \cite{Shoval_Sheftel_Shinar_Hart_Ramote_Mayo_Dekel_Kavanagh_Alon_2012}. In the periphery of sensory systems, like that of the blowfly's visual pathway, for example, maximization of information capacity seems to be the central goal \cite{Laughlin_1981}. Any information not captured at this stage is lost to the system. Further downstream neurons, however, are thought to more strongly adhere to additional objectives, like energetic or structural cost, resulting in more efficient, yet narrowed-down representations of behaviorally relevant features \cite{Clemen2011EfficientTrans}. Often a compromise to apparent incompatibilities in objectives is required, and Pareto theory, as discussed in this review, allows to indentify the least Salomonic option.

\section*{Pareto Optimality}

\begin{figure}[H]
    \centering
    \includegraphics[width=\textwidth]{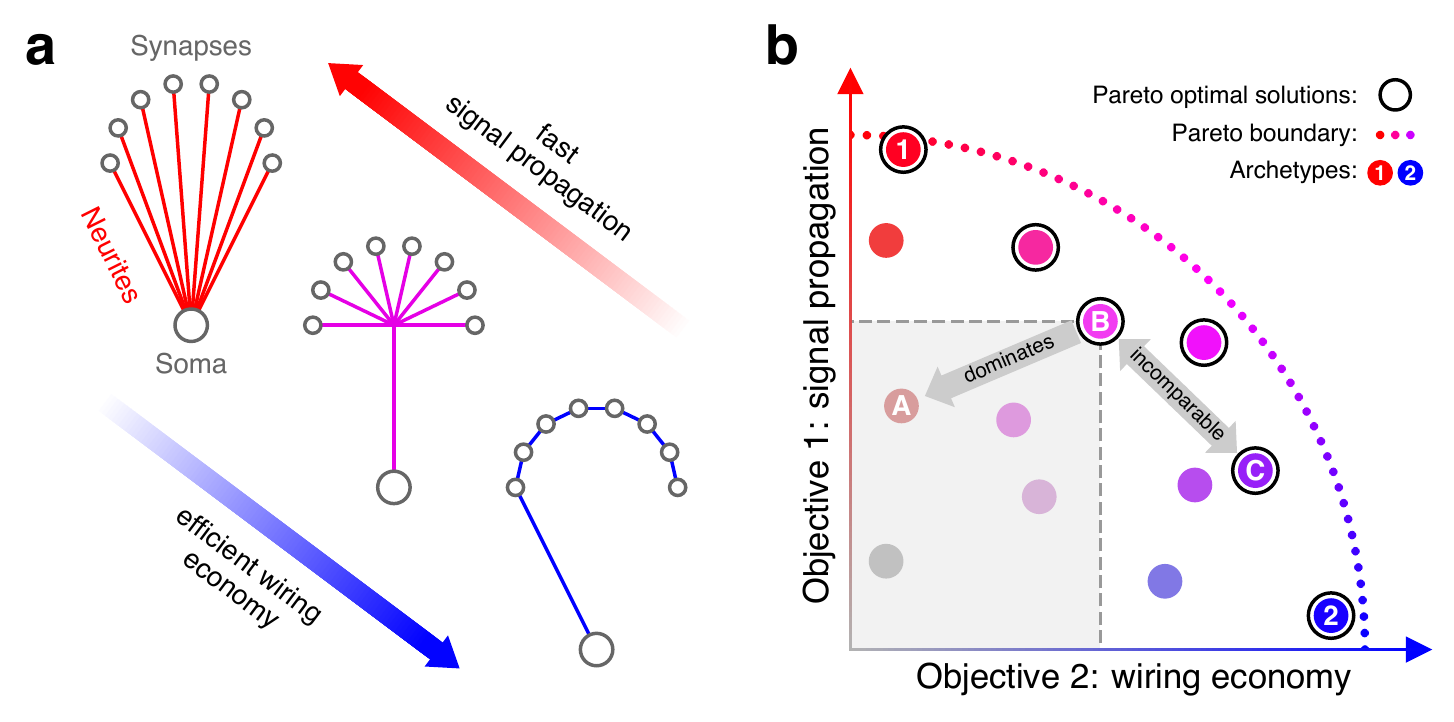}
    \caption{\small Pareto optimality, visualized using an example of neural arbors that aim to maximize two conflicting objectives: fast signal propagation and efficient wiring economy \cite{chandrasekhar_neural_2019}. (\textbf{a}) Three arbors with different configurations of neurites (axons/dendrites) that connect a fixed set of synapses to a soma. The arbor that connects each synapse to the soma with a separate neurite (top left) minimizes conduction delays. It is the optimal solution to (and thus constitutes the archetype for) fast signal propagation, but comes at a high wiring cost. Vice versa, the tree that optimizes wiring economy (bottom right) inevitably increases conduction delays. The arbor in the middle constitutes a compromise in the trade-of between the two objectives. (\textbf{b}) Some possible solutions in the space of the two objectives. Solution B, for example, dominates A. The gray rectangle marks the space of all solutions that are dominated by B. Solutions C and B are both superior to each other in at least one objective and therefore incomparable. Solutions that are not themselves dominated (circled dots) are considered Pareto optimal. The set of all Pareto optimal solutions ("Pareto set") approximates the Pareto boundary, here represented as an idealized curve (dotted line). The digits 1 and 2 mark the archetypes, i.e. the solutions that maximize objective 1 and 2, respectively.}
    \label{fig:pareto}
\end{figure}

Conflicting objectives are apparent in nervous system function and structure, as illustrated in the following example (Figure \ref{fig:pareto}). Neuronal arborization is determined by the trade-off between two objectives \cite{chandrasekhar_neural_2019}: wiring economy and propagation speed. The former dictates the minimization of total arbor length because of its structural energetic cost; the latter, however, improves with shorter path lengths between a neuron's soma and synapses (Figure \ref{fig:pareto}a). In such cases, where multiple objectives compete, the concept of Pareto optimality enables us to differentiate objectives and even to predict their relative evolutionary impact from data.

According to Pareto theory, a solution is \emph{Pareto optimal} if no objective can be improved without simultaneously worsening others. In our example, a neural arbor is considered Pareto optimal if no other arbor exists that is better at maximizing either economic wiring or speed without decreasing the other (Figure \ref{fig:pareto}b). Conversely, if another arbor existed that was better in one objective without compromising in the other, it would \emph{Pareto dominate} the original arbor (and the latter would, consequently, not be considered Pareto optimal). From these definitions it also follows that multiple neural arbors can be Pareto optimal. 
If one arbor has a better wiring economy while another has faster signal propagation, the two morphologies are considered \emph{Pareto incomparable}, meaning neither solution is dominating the other. The favorable solution is then determined by additional constraints, like explicit assumptions on the weighting between these objectives (cost versus speed) or the inclusion of an additional objective (like size limitations).  
In principle, all Pareto-optimal solutions are incomparable. They form the Pareto set, in our example comprising all neural arbors that are not dominated by other morphologies. The Pareto set approximates the idealized \emph{Pareto boundary}, and represents alternative optimal solutions which depend on the weighting between different objectives (Figure \ref{fig:pareto}b). For insights on evolutionary principles, it is particularly worthwhile to compare where a biological system (like an observed arbor) is positioned relative to the Pareto boundary (obtained from the exploration of computer-generated model arbors); both proximity to and specific location on the boundary provide relevant information.

\section*{Network Wiring Respects Both Wiring Cost and Propagation Efficiency}
 Chandrasekhar \& Navlakha \citep{chandrasekhar_neural_2019} not only showed that neural arbors across multiple brain regions and species lie close to the Pareto boundary regarding conduction delays and total wire length, but where neurons fall on this boundary is not random. Different neuron types weigh wiring cost and conduction delays differently, with cells involved in tasks that need to be executed quickly presumably assigning more weight to minimizing conduction delays. 
If attention is only paid to wiring length, better solutions are often theoretically possible (see examples in macaques and \textit{C. elegans} \cite{ Kaiser_Hilgetag_2006,Avena-Koenigsberger2014, Gushchin_Tang_2015}). Still, biological networks perform significantly better than random networks - take the diffuse motor net of a jellyfish as an example, where wiring economy is improved via the orientation of its neurites \cite{Pallasdies_Goedeke_Braun_Memmesheimer_2019}. While neurons seem to be optimizing for wiring length, a multi-objective optimization approach is required to gain a holistic understanding of the underlying evolutionary strategies. In fact, other systems like plant shoots and even city roads appear to optimize the same objectives of structural cost and speed, supporting the idea of an overarching principle guiding the generation of such structures \cite{Suen_Navlakha_2019}.

Similar results can be found in large scale neural structures and connectomics data. When a Nash equilibrium network game model -- which maximizes navigability while minimizing wiring cost \cite{Gulyas2015} -- is applied to the human connectome, the structural connectivity closely resembles the model network \cite{Pappas_Craig_Menon_Stamatakis_2020}. It was also found that this optimal connectivity predicts resting state functional connectivity but not task-based functional connectivity, presumably because more demanding tasks recruit non-optimal connections that allow communication between different regions and support integration at the whole-brain level \cite{Betzel_Bassett_2018}.


\section*{Trade-offs Involving Limited Energetic Supplies}

The need to efficiently use limited energetic resources is fundamental for neural systems. This objective has been extensively studied, from ion-channel-based mechanisms to morphology \cite{Attwel2001AnEnergyBudge,Schrei2002EnergyEfficien,Sengup2010ActionPotentia,Hasenstaub_Otte_Callaway_Sejnowski_2010,Sengupta2014Power,Roemschied2014Cell,Hesse_Schreiber_2015, Heras_Anderson_Laughlin_Niven_2017}. Often energetic cost and functionality are interrelated. Spiking costs, for instance, vary with frequency and firing pattern \cite{Yi_Wang_Tsang_Wei_Deng_2015, Yi_Grill_2019,Yi_Wang_Wei_Che_2019}, can be influenced by spike frequency adaptation \cite{Yi_Wang_Li_Wei_Deng_2016}, or EPSP timecourse \cite{Harris_Jolivet_Engl_Attwell_2015, Yuan_Huo_Fang_2018}. In particular when spike rates are high, interesting trade-offs can reveal themselves, as the following two examples show. 

\begin{figure}[H]
    \centering
    \includegraphics[width=1\textwidth]{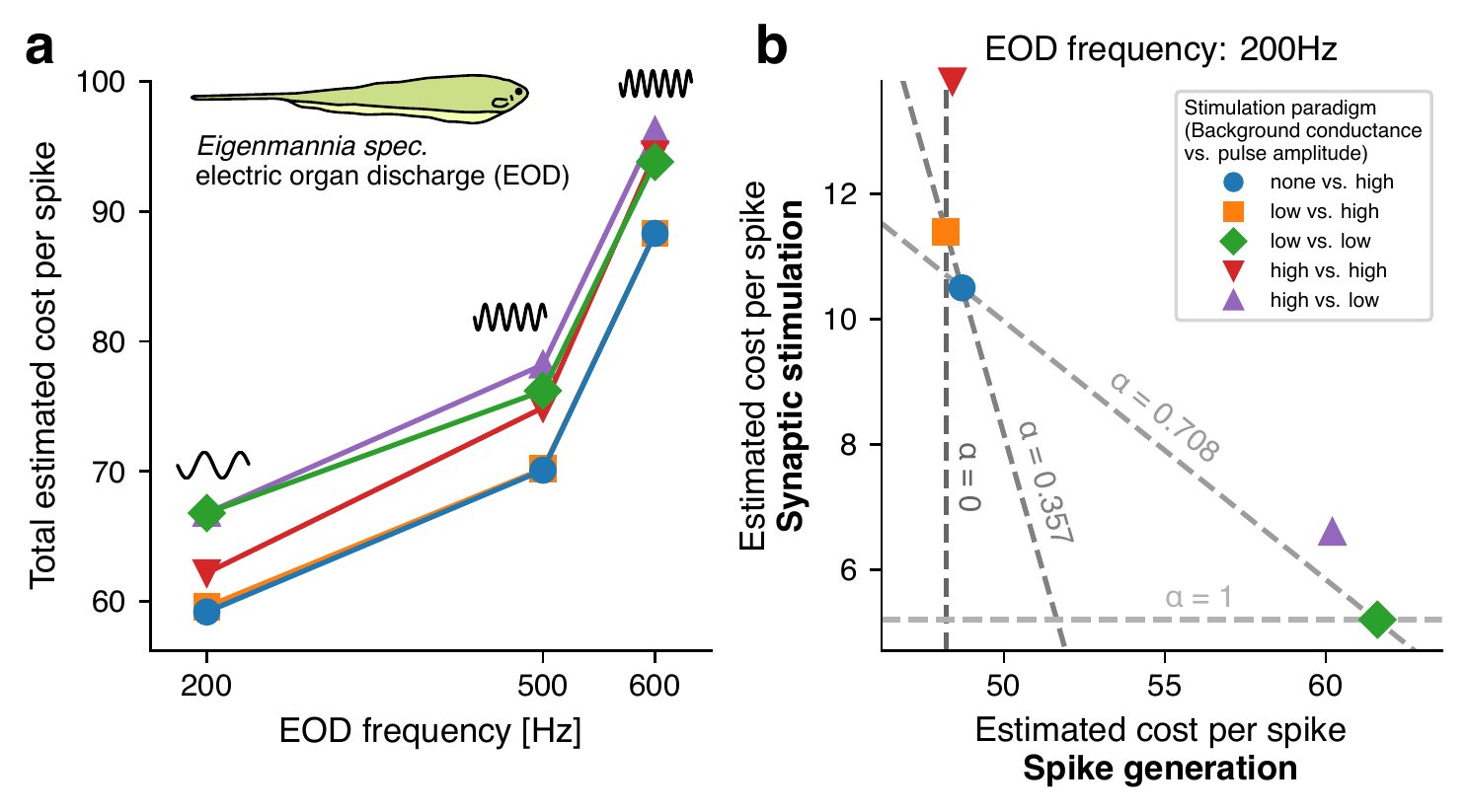}
    \caption{\small Energetic cost of sustained high-frequency firing in a model of the weakly electric fish \textit{Eigenmannia} electric organ (data from \cite{Joos_Markham_Lewis_Morris_2018}), stimulated via postsynaptic acetylcholine receptor (AChR) activation (regular pulse-trains with a constant background conductance). (\textbf{a}) Resulting energy consumption of the electric organ discharge (EOD) estimated from the total entry of sodium ions per action potential into the cell (Na$^+$-entry/AP $*10^9$, y-axis). Shown are the results of five simulations across different pulse train frequencies (for five combinations of pulse amplitude and background conductance).
    (\textbf{b}) The same five simulations for an EOD frequency of 200Hz, with total estimated cost split into two separate objectives: cost of spike generation (x-axis) and cost of synaptic stimulation (y-axis) -- corresponding to Na$^+$ entry through voltage-gated Na$^+$ channels and entry through AChRs. Synaptic stimulation is likely to contribute more to overall energetic cost, as it also requires neurotransmitter release and re-uptake. A weighting parameter $\alpha$ allows to determine the optimal solution $S$, which minimizes the Pareto cost $\alpha O_1(S) + (1 - \alpha) O_2(S)$ of the objectives $O_1$ and $O_2$ for a given value of $\alpha$. Lines of equal Pareto cost for different values of $\alpha$ are shown in gray. $\alpha = 0$ and $\alpha = 1$ are the extremes, of which the optimal solutions are the archetypes of the respective objectives. Also shown are intermediate values of $\alpha$ where two solutions have equal Pareto cost ($\alpha = $ 0.357 and 0.708). For example, if the overall metabolic cost of Na$^+$ entering through AChRs is more than 1/0.708 = 1.41 times as high as Na$^+$ entering through voltage-gated Na$^+$ channels, the simulation represented by the green diamond becomes optimal.}
    \label{fig:joos}
\end{figure}

\textbf{Electric organ discharge in the weakly electric fish:} The genus \textit{Eigenmannia} uses a specialized electric organ to produce discharges at a near-constant rate between 200 and 600 Hz, both for electrosensing and communication. Based on biophysical models, Joos and colleagues \cite{Joos_Markham_Lewis_Morris_2018} showed that the energetic cost per spike, estimated from the sodium current flow, increases superlinearly with firing rate (Figure \ref{fig:joos}a). Na$^+$ expulsion through the Na$^+$/K$^+$ pump constitutes a large part of the energy demand of spiking \cite{Laughlin_de_1998}, yet the quantification of metabolic cost is complex \cite{Hasenstaub_Otte_Callaway_Sejnowski_2010, Moujahid_dAnjou_Torrealdea_Torrealdea_2011} and other processes contribute \cite{Engl_Attwell_2015}. For example, Na$^+$ entry via postsynaptic receptor channels incurs additional costs through neurotransmitter packaging, release and uptake. The relative contribution of different costs and their impact on optimal solutions can be nicely explored in the Pareto framework, as we propose in Figure \ref{fig:joos}.

\textbf{Coincidence detection in the Medial Superior Olive (MSO):} Pareto analysis has been successfully applied in the MSO \cite{Remme_Rinzel_Schreiber_2018}. A systematic analysis of morphological and electrophysiological properties (like ion-channel densities and dendrite and soma sizes) in mathematical models, showed that MSO cells are indeed located closely to the Pareto-optimal set. Placing the experimentally-determined characteristics of MSO cells in the physiological parameter space evidenced their operation in a regime where functional performance is almost optimal and energy savings have only impacted neuronal design where performance is not compromised (a "function first, energy second" principle). It appears that oftentimes, neurons are optimized for objectives not directly related to functional performance until the law of diminishing returns sets in, where a small further improvement in the objective would strongly decrease the functional abilities (see also \cite{Bryman_Liu_Do_2020} for an example supporting this observation in primate photoreceptors).

\section*{Trade-offs in the Neural Code}

To complement the picture, also the neural code is subject to trade-offs \cite{Sengupta_Laughlin_Niven_2013,Yu_Zhang_Liu_Yu_2016, Marzen_DeDeo_2017}, which we can only touch on here by way of example. Stringer et al. \citep{Stringer_Pachitariu_Steinmetz_Carandini_Harris_2019} imaged neural activity in the visual cortex and demonstrated that neurons were neither in an \textit{efficient} coding regime, where activity is uncorrelated and high-dimensional, nor in a \textit{robust} regime, where activity is low-dimensional and correlated, i.e. redundant. Activity was found to be high-dimensional, yet the correlation closely followed a specific power-law. The authors argue that if neurons were any less correlated, the population code would be compromised, with similar input images not resulting in similar output activities. This trade-off between robustness and efficiency seems to be universal, but the weighting can be different between brain areas and species. For example, human cingulate cortex and amygdala have been found to be more efficient but less robust than these areas in macaques \cite{Pryluk_Kfir_Gelbard-Sagiv_Fried_Paz_2019}.
Also coding of networks in critical regimes can cluster along the Pareto boundary, implementing compromises between information propagation and metabolic cost \cite{Aguilar-Velazquez_2021}. Generally, trade-offs in coding also extend to behavior and learning. An organism should be able to quickly adapt to environmental changes while being robust enough not to be affected by noise \cite{Raman_Rotondo_OLeary_2019, Nassar_Troiani_2020, Duggins_Krzeminski_Eliasmith_Wichary_2020} or not to overwrite established memories. The latter constitutes the classical plasticity-stability dilemma \cite{Abraham_Robins_2005, Mermillod_Bugaiska_Bonin_2013, Verbeke_Verguts_2019}, which is a special case of Pareto optimality \cite{jin2006alleviating}. We note that for behavioral paradigms, data-driven Pareto approaches are often more fruitful than the model-based ones.

\section*{A Statistical Approach to Pareto Theory: Examples from Behavior}
Above, Pareto theory has been combined with a model-based approach, enabling the computational study of trade-offs that yields a full description of the optimal parameter space. 
In highly complex systems, however, accurate mechanistic models may not be available. Alternatively, a statistical approach to Pareto theory can be taken. Here, correlations within more elaborate datasets of measured experimental 'traits', for example, across species in different habitats or of the same animal across different behavioural states, can be exploited to estimate relevant objectives that are consistent with the observed correlations. The approach is data-driven, no mechanistic assumptions are required.

The statistical analysis starts from an experimental data set that characterizes traits across a larger population. 
The term 'traits' refers to a number of characteristics, from morphological measurements and gene expression profiles to behavioral statistics, that are suspected to be relevant for the potential optimization under study. The aim is to find out whether the distribution of traits in the data can be explained by trade-offs between multiple objectives and, ideally, what these objectives are (without prior assumptions on their nature).

When the data is plotted as a function of the individual traits, one expects certain correlations in them, if they lie on a Pareto boundary. Namely, a specific geometrical shape, a so-called polytope can be found, that covers the data in this space of traits \cite{Shoval_Sheftel_Shinar_Hart_Ramote_Mayo_Dekel_Kavanagh_Alon_2012}. This relationship is explained further in the box.

The vertices of these structures correspond to archetypes (where optimality predominantly arises from one objective), their number indicating the number of relevant objectives. Data points lying within these geometrical structures define Pareto-optimal solutions, each point representing a different weighting between the objectives (see Fig. \ref{fig:morphospace}) . For the detection of polytope shapes, statistical methods like t-ratio tests are used. Their application, however, is non-trivial due to a high false-discovery rate in phylogenetic data (see Box). 

To illustrate the approach, we turn towards the analysis of locomotive states in \textit{C. elegans}. Gallagher et al. \cite{Gallagher_Bjorness_Greene_You_Avery_2013} measured the animal's movement direction and speed, observing different behaviours. Previously, these had been described as stereotyped, but when the data was analyzed, a continuum of states was identified. Represented in the two-dimensional trait space (defined by direction and speed measurements), 
a triangle is revealed (see Fig. \ref{fig:morphospace}b for a schematic). This geometry suggests the relevance of \textit{three} objectives. Triangle vertices represent the behavioural archetypes; for \textit{C. elegans} these have been identified as roaming, dwelling and quiescence. In terms of the objectives associated with each archetype, the interpretation as exploration, exploitation, and energy conservation suggests itself. Variations in locomotive behavior (i.e. deviations from the archetypes) can be explained by behaviours that evolved along the Pareto boundary (here formed by the whole triangle), data points within the triangle representing trade-offs between the objectives according to the surrounding environment and the organism's current needs.

This method has also been applied to behavioral data in other animals, including humans, to identify behavioral archetypes \cite{Forkosh_Karamihalev_Roeh_Alon_Anpilov_Touma_Nussbaumer_Flachskamm_Kaplick_Shemesh_2019,Cona_Kocillari_Palombit_Bertoldo_Maritan_Corbetta_2019}.
It has also been used on human imaging data of lateralized activity patterns \cite{Karolis_Corbetta_ThiebautdeSchotten_2019}. These are brain activity patterns during behavioral tasks that are asymmetric between the two hemispheres. By studying this task evoked activity (with traits defined by a low-dimensional representation of the activity) one finds brain functional lateralization is structured in a polytope with four vertices suggesting four optimized objectives, which have been interpreted as symbolic communication, perception, emotional contextualization and decision-making.

\begin{center}
\begin{singlespace}
\fcolorbox{darkgray}{Gainsboro}{\begin{minipage}{\textwidth}
\textbf{\Large Statistical Inference of Pareto Optimality}
\small

The statistical inference of Pareto optimality relies on the assumption that each objective function is maximized by a single \emph{archetype} (numbered data points in Figure \ref{fig:morphospace}) and decreases solely with distance in the space of traits (see radial contours around the archetypes). In essence this means that the more a given phenotype differs from an archetype, the poorer its performance in that objective is. Pareto theory then predicts all optimized phenotypes should fall into a polytope spanned by the archetypes (line, triangle, tetrahedron) \cite{Shoval_Sheftel_Shinar_Hart_Ramote_Mayo_Dekel_Kavanagh_Alon_2012, Hart_Sheftel_Hausser_Szekely_Ben-Moshe_Korem_Tendler_Mayo_Alon_2015}. 
For two objectives (Figure \ref{fig:morphospace}a) this can be understood by plotting the combined objective along different trajectories connecting the two archetypes. As long as the objective fitness decreases uniformly in all directions with distance from the archetypes (i.e. isotropically) the shortest pathlength connection (the line) is optimal (Fig. \ref{fig:morphospace}a inset). For any point that is deviating from this line there will be a point on it (namely its projection onto the line), that is closer to both archetypes, and since the objectives decrease with distance from these archetypes, the latter point would perform better in both objectives and therefore Pareto dominate the former. This could underlie the many linear correlations between traits seen in nature. 

Depending on the number of objectives and archetypes, datapoints will lie on a line segment for two objectives, on a triangle for three objectives (Fig. \ref{fig:morphospace}b), and on higher-dimensional polytopes for a higher number of objectives.
If the assumptions above are relaxed a bit and the objective function does not decrease uniformly in all directions from the archetypes, the resulting contour plot in Fig. \ref{fig:morphospace} would not consist of concentric circles (i.e., objectives are nonisotropic)  the line along which the Pareto optimal solutions fall would be more complex, which in statistics is known as the ridgeline manifold \cite{RayS2005TheTopography, Shoval_Sheftel_Shinar_Hart_Ramote_Mayo_Dekel_Kavanagh_Alon_2012}.

Statistical tests for Pareto optimality make use of the above geometric observation under the assumption of isotropic objectives. Unfortunately, there are several mechanistic and phylogenetic reasons for trait correlations unrelated to Pareto optimality. Specifically, fitted polytopes to phenotypic data sets have been tested for significance based on a $t$-ratio test \cite{Mikami_Iwasaki_2021}; the fit to the polytope is compared with a shuffled version to estimate significance. Shuffling the data, however, assumes that traits are uncorrelated and the method, therefore, tends to overestimate significance \cite{SunM2020RampantFalseD, Mikami_Iwasaki_2021}.

\begin{figure}[H]
    \centering
    \includegraphics[width=0.95\textwidth]{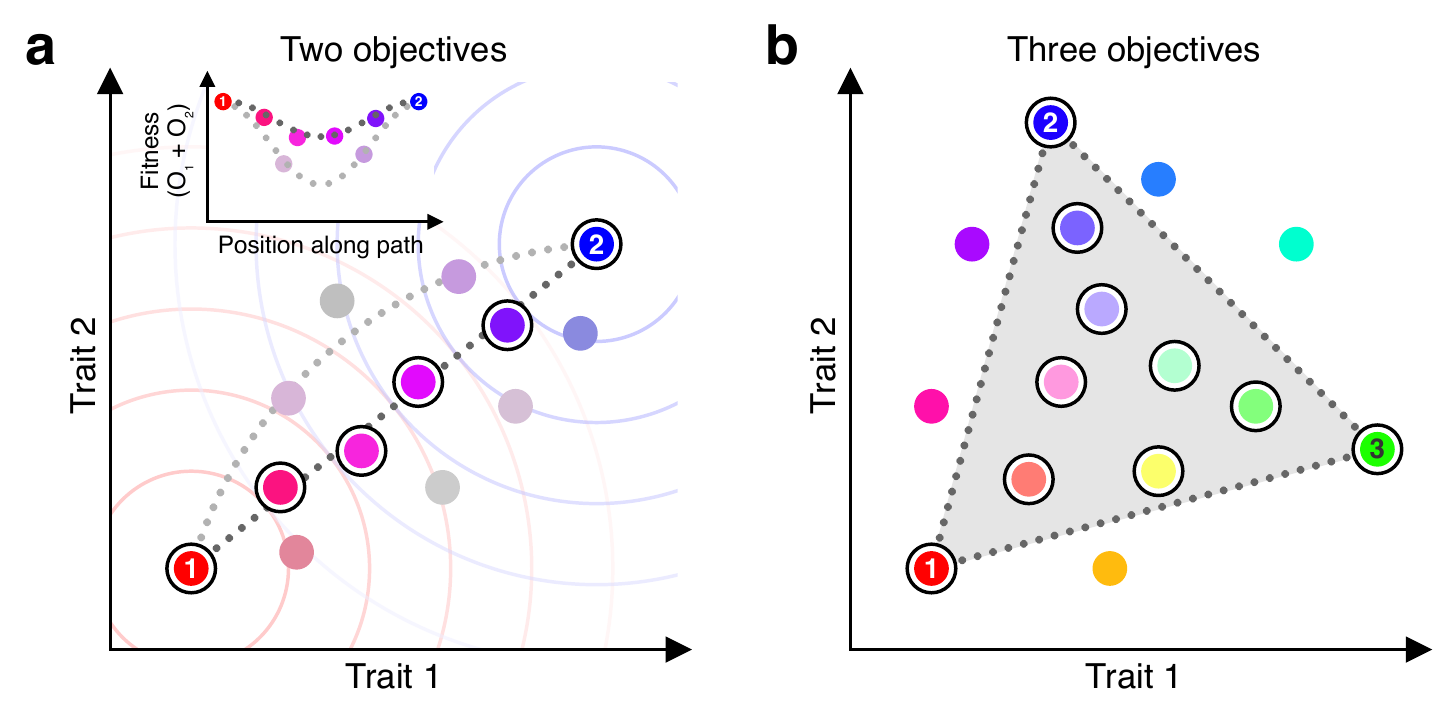}
    \caption{\small Pareto optimality in a 2-dimensional traitspace. (\textbf{a}) In traitspace, the Pareto boundary of two objectives forms a straight line between the two archetypes (numbered dots) under certain assumptions \cite{Shoval_Sheftel_Shinar_Hart_Ramote_Mayo_Dekel_Kavanagh_Alon_2012}. Pareto optimal solutions (circled dots) appear along the line. The closer a solution is to one of the archetypes, the higher the weight that is placed on the corresponding objective. Radial contours indicate isotropic objective functions $O_1$ and $O_2$ that decrease with distance from the archetypes. Their weighted sum is maximized along the Pareto boundary (dark dotted line). The inset shows one example of such a weighting of the objectives and the combined fitness of solutions along different paths between the archetypes. Note that the highest combined fitness does not need to lie at the archetypes.  (\textbf{b}) Three objectives form a triangle in traitspace, higher numbers of objectives form polytopes, whose vertices are the archetypes of each objective (1--3). Optimal solutions (circled dots) are expected to fall within the polytope space spanned by the archetypes. Traitspace can consist of an arbitrary number of dimensions.}
    \label{fig:morphospace}
\end{figure}
\end{minipage}}
\end{singlespace}
\end{center}


\section*{Conclusion}

Trade-offs (on multiple scales) are integral to neural organization, yet the relevant objectives are not always easy to identify. If a Pareto analysis reveals that evolved structures do not lie on a model-derived Pareto boundary, this may either be the case because they are not optimal or, alternatively, because relevant objectives were missing in the analysis. In the latter case, novel hypotheses about relevant design principles of the investigated system can be formulated and put to test.

Moreover, the Pareto approach can help to decipher generic principles of nervous system evolution. It is not unlikely that some nervous systems may have evolved along Pareto boundaries, accompanied by a gradual change in the weighting between objectives over time. For example, when a new structure evolved, it may at first have only sub-optimally fulfilled its new function, favouring those structures whose energetic demand was sufficiently low not to impair overall fitness. 
However, the more the function improved and benefited fitness, the more the trade-off may have gradually shifted along the Pareto boundary from a higher impact of energetic constraints towards an increased relevance of functional performance, ultimately resulting in a "function first, energy second" archetype. 

In contrast to other fields, like economy, Pareto optimality has only recently been applied to neuroscientific problems, particularly in the realm of computational neuroscience, where the space of possible solutions can be explored through mechanistic modelling approaches. More statistical tools are needed to also identify Pareto boundaries and archetypes directly from experimental data. Both model- and data-driven Pareto approaches have great potential to advance our understanding of the principles that guide the evolution of neural structures.


\section*{Acknowledgments}

This work was supported by the German Ministry of Education and Research (grant no. 01GQ1403), the Deutsche Forschungsgemeinschaft (DFG, German Research Foundation – Project number 327654276 – SFB 1315) and the European Research Council (grant no. 864243).


\begin{center}
\begin{spacing}{1.3}
\fcolorbox{darkgray}{Gainsboro}{\begin{minipage}{\linewidth}
\textbf{\Large Outstanding Papers}
\begin{itemize}
    \small
    \item \cite{DelGiudice_Crespi_2018} * The authors categorize the distributed literature on neural trade-offs and build the foundation for an overarching framework.
    \item \cite{chandrasekhar_neural_2019}** The authors are able to show that neural arbors are close to the Pareto boundary of wiring length and transduction delay and that different neuron types weigh these objectives differently.
    \item \cite{Bryman_Liu_Do_2020} ** Through computational modelling the authors show that primate photoreceptors are as long and narrow as possible without a loss on signal transduction to allow for a high resolution near the fovea.
    \item \cite{Pappas_Craig_Menon_Stamatakis_2020} * A Nash equilibrium network game model that maximizes navigability and minimizes wiring cost predicts resting-state connectivity but not functional connectivity during working memory tasks.
    \item \cite{Joos_Markham_Lewis_Morris_2018} ** Using a computational model of the electric fish's organ discharge, the authors showed that with an increase in frequency, the energetic costs of APs rise superlinearly while the synaptic costs remain constant. This potentially predicts channel densities across fish.
    
    \item \cite{Remme_Rinzel_Schreiber_2018} ** By systematic analysis of morphological and electrophysiological properties, it is shown that the principal cells of the medial superior olive are located near the Pareto boundary with regards to their energy efficiency and functional performance in coincidence detection.
    
    \item \cite{Stringer_Pachitariu_Steinmetz_Carandini_Harris_2019} ** By recording an astounding 10,000 neurons in the visual cortex, the authors were able to show that the neuron activities are just correlated enough for the coding to be as efficient as possible, while still producing a smooth representation.
    \item \cite{Pryluk_Kfir_Gelbard-Sagiv_Fried_Paz_2019} * Through careful construction of a measure for efficiency and robustness, different brain areas are shown to lie at different places on this trade-off and that human brain areas are more efficient but less robust than the same areas in macaques.
    \item \cite{Raman_Rotondo_OLeary_2019} * Through an elegant modelling approach, the authors show that a learning network with a given level of intrinsic noise has an optimal network size to learn as quickly as possible.
    \item \cite{Forkosh_Karamihalev_Roeh_Alon_Anpilov_Touma_Nussbaumer_Flachskamm_Kaplick_Shemesh_2019} *  By decomposing the high-dimensional behaviors of mice in social boxes, the authors found four identity domains that correlate with standard behavioral tests and tanscriptomic variances in the brain. Through Pareto task inference, the two strongest domains span a  triangle determined by three archetypes.
    
    \item \cite{Mikami_Iwasaki_2021} **The authors show that the standard test for Pareto optimality in trait space can have a high false discovery rate due to phylogenetic signals and propose an alternative test, that addresses this issue.

\end{itemize}
\end{minipage}}
\end{spacing}
\end{center}
\vspace{1cm}
\bibliographystyle{vancouver}

\end{document}